\documentclass[smallextended]{svjour3}

\smartqed

\usepackage[T1]{fontenc}
\usepackage[utf8]{inputenc}
\usepackage{lmodern}

\usepackage{amsmath,amssymb,mathtools,bm}
\usepackage{graphicx}
\usepackage{physics}
\usepackage{booktabs}
\usepackage{siunitx}

\usepackage{hyperref}

\begin{document}

\title{Constraint-Limited Tube Orientation of Entangled Polymers in Oscillatory Shear Deformation}

\author{
Dario Nichetti
\and
Alessio Zaccone
}

\institute{
Dario Nichetti \at
Rheonic Lab, Via Quadelle 2C, 26012 Castelleone (CR), Italy
\email{dario.nichetti@rheoniclab.com}
\and
Alessio Zaccone \at
Department of Physics ``Aldo Pontremoli'',
University of Milan,
Via Celoria 16,
20133 Milan,
Italy
\email{alessio.zaccone@unimi.it}
}

\date{Received: date / Accepted: date}

\maketitle

\begin{abstract}
We develop a molecularly motivated description of the nonlinear index (NLI) in oscillatory shear deformation of entangled polymers. The central assumption is that the shear component of the tube-orientation tensor cannot grow without bound. Convective constraint release (CCR), chain stretch, and tube dilation progressively reduce the number and lifetime of orientational constraints, but the maximum shear alignment of a tube segment is geometrically limited by $S_{xy}\leq 1/2$. This motivates a constraint-limited orientation closure in which the NLI first grows approximately with strain amplitude and then approaches the limiting value $\mathrm{NLI}_{\max}=3$ asymptotically rather than through an artificial cutoff.
The same framework yields a molecular expression for the characteristic half-saturation strain $\gamma_s$, defined by $\mathrm{NLI}(\gamma_s)=3/2$, in terms of the entanglement number, oscillation frequency, and a critical number of remaining orientational constraints. We further derive architecture-dependent expressions for the nonlinear onset strain $\gamma_c$ for linear, sparsely long-chain-branched, and more regularly branched polymers.
The resulting framework provides a compact bridge between Fourier harmonic analysis, CCR-based tube dynamics, and the progressive loss of orientational memory in highly deformed entangled polymer liquids.

\end{abstract}

\section{Introduction}
Periodic shear provides a sensitive framework for probing the evolution of polymer elasticity from the linear viscoelastic regime to strongly nonlinear deformation. At small strain amplitude, the stress response is sinusoidal and dominated by the first harmonic. As the strain amplitude is increased, the stress progressively develops higher harmonics that encode the molecular relaxation mechanisms activated within the entangled tube network \cite{Hyun2011}.

The use of oscillatory shear as a structural probe is justified because nonlinear deformation converts otherwise hidden molecular relaxation mechanisms into measurable higher harmonics, which can be analyzed through Fourier-transform rheology \cite{Wilhelm1998,Wilhelm2002} or through time-domain nonlinear material measures \cite{Ewoldt2008}. For entangled polymer melts and concentrated solutions, these nonlinear harmonics reflect the deformation of a transient tube network whose orientational memory is progressively modified by reptation, chain stretch, tube dilation, and convective constraint release (CCR) \cite{deGennes1971,DoiEdwards1986,RubinsteinColby2003,Marrucci1996,IannirubertoMarrucci1996}.

The linear tube model already shows that reptation alone is not sufficient to describe entangled-chain relaxation, because contour-length fluctuations, longitudinal stress relaxation, and constraint release must be included to obtain realistic relaxation spectra \cite{LikhtmanMcLeish2002}. At larger deformation rates, the same molecular picture must be extended to include deformation-driven constraint renewal and chain stretch, because orientation and stretch relax on different molecular time scales \cite{Milner2001,Graham2003}. Modern tube models incorporate CCR together with chain stretch and constraint renewal as essential ingredients for describing nonlinear flows of entangled polymers over a broad range of deformation rates \cite{LikhtmanGraham2003}.

At a more microscopic level, in disordered materials, deformation is generally not purely affine at the molecular scale. Under an affine deformation, every particle, segment, or molecular unit follows exactly the macroscopic strain field. In contrast, structural disorder breaks local inversion symmetry and generates unbalanced forces under deformation, causing additional molecular displacements that restore local mechanical equilibrium. These extra relaxational motions are termed \emph{nonaffine displacements} and reduce the elastic energy stored by the system~\cite{ZacconeScossaRomano2011,ZacconeBook2023}. In amorphous solids and polymer glasses, the elastic modulus can therefore be decomposed into an affine contribution and a negative nonaffine correction arising from these molecular relaxations~\cite{ZacconeScossaRomano2011,ZacconeTerentjev2013}. From a molecular perspective, nonaffinity quantifies the extent to which local rearrangements, constraint release, and configurational relaxation prevent the material from following a perfectly homogeneous deformation field. Although entangled polymer melts differ from polymer glasses in that their constraints are topological rather than permanent, the same physical concept remains useful: nonlinear flow progressively activates relaxation mechanisms that reduce the coherent load-bearing response of the tube network. In this sense, convective constraint release (CCR), tube renewal, and loss of orientational memory may be viewed as dynamic manifestations of nonaffine relaxation in entangled polymers.
In this framework, the stress waveform is not merely a distorted sinusoid but a compact projection of intra-cycle elasticity, strain softening, and the progressive loss of proportionality between stress and strain \cite{Cho2005,EwoldtMcKinley2010,Ewoldt2007}.

Recently, the Nonlinearity Index (NLI) was introduced as a quantitative measure of nonlinear viscoelasticity based on the relative magnitude of the higher elastic harmonics generated during LAOS deformation~\cite{NichettiScacchi2025,ScacchiNichetti2025AJOP,NichettiScacchi2025KGK}. The NLI is defined from the ratio between the nonlinear elastic contribution reconstructed from the odd higher harmonics and the first-harmonic storage modulus. Experimentally, it provides a sensitive fingerprint of nonlinear molecular relaxation and has been shown to collapse data from a wide range of polymeric systems onto common master curves. In a previous nonaffine interpretation of LAOS, the nonlinear elastic harmonics were associated with the transfer of elastic response from coherent first-harmonic storage into nonlinear relaxation channels~\cite{nichetti2026}. The present manuscript develops a complementary molecular closure for this idea. The key point is simple: once the primitive-path segments have approached their maximum shear alignment, further imposed strain cannot be stored as proportional additional tube orientation. Instead, additional deformation is mainly dissipated through faster constraint renewal, CCR, and tube relaxation.

The present NLI definition goes beyond conventional third-harmonic measures because it collects all higher odd elastic harmonics into a single normalized quantity while preserving the physical role of the first harmonic as the coherent storage response. In this sense, the NLI provides a compact scalar measure of the cumulative transfer of elastic response into nonlinear modes \cite{Neidhofer2003,HyunWilhelm2009,Hoyle2014}.

The theory is built around the following related questions. First, what sets the maximum admissible value of the nonlinear index (NLI)? Second, how should the linear or approximately proportional NLI growth, observed in the nonlinear regime, bend toward this limiting value? 

The answer proposed here is a constraint-limited tube-orientation regime. The tube remains dynamic at high strain amplitude; it is not frozen. However, its capacity to store additional shear orientation is exhausted by the combined action of chain stretch, tube dilation, and CCR. This produces an asymptotic NLI curve and a characteristic strain for the onset of nonlinearity.

\section{Harmonic decomposition and definition of the NLI}

We consider sinusoidal shear
\begin{equation}
    \gamma(t)=\gamma_0\sin(\Omega t),
\end{equation}
where $\gamma_0$ is the strain amplitude and $\Omega$ is the angular frequency. In the nonlinear regime, the elastic stress contains odd sine harmonics,
\begin{equation}
    \sigma_E(t)=\sum_{k=1,3,5,\ldots}\sigma'_k(\gamma_0,\Omega)\sin(k\Omega t).
\end{equation}
The first-harmonic storage modulus is
\begin{equation}
    G'_1(\gamma_0,\Omega)=\frac{\sigma'_1(\gamma_0,\Omega)}{\gamma_0}.
\end{equation}
Following the generalized nonlinear elastic modulus construction used in NLI analysis, we define
\begin{equation}
    G'_{NL}(\gamma_0,\Omega)=\sum_{n=1}^{\infty}(2n+1)\frac{\sigma'_{2n+1}(\gamma_0,\Omega)}{\gamma_0}.
\end{equation}
The nonlinear index is then
\begin{equation}
\label{eq:NLI_definition}
    \mathrm{NLI}(\gamma_0,\Omega)=-\frac{G'_{NL}(\gamma_0,\Omega)}{G'_1(\gamma_0,\Omega)}.
\end{equation}

The sign convention in \eqref{eq:NLI_definition} is appropriate for strain-softening nonlinear elasticity, where the higher-harmonic nonlinear correction is negative while the reported NLI is positive. The above separation between the first-harmonic elastic response and the nonlinear harmonic correction is also consistent with broader classifications of rheological nonlinearity, in which nonlinear behavior is identified by the failure of a single amplitude-independent material function to represent the stress response \cite{Giesekus1982,Malkin1995}.

\section{Maximum admissible shear orientation}

Let $\bm{u}$ be the unit tangent vector along a tube segment or primitive-path segment. The shear component of the tube-orientation tensor is
\begin{equation}
    S_{xy}=\langle u_xu_y\rangle .
\end{equation}
Since $\bm{u}$ is a unit vector, the maximum possible value of $u_xu_y$ is obtained when the segment lies in the shear plane and is oriented at $45^\circ$:
\begin{equation}
    u_x=u_y=\frac{1}{\sqrt{2}},\qquad u_z=0.
\end{equation}
Therefore,
\begin{equation}
\label{eq:Sxymax}
    S_{xy}^{\max}=\frac{1}{2}.
\end{equation}

In the constraint-limited orientation picture adopted here, the nonlinear elastic distortion is assumed to scale proportionally with the shear-orientation amplitude. Thus, within the present closure, the nonlinear index is interpreted as a Fourier-resolved measure of how much elastic response has been transferred from coherent first-harmonic storage into nonlinear orientational modes. Since tube-based descriptions relate polymer stress directly to the orientational state of primitive-path segments, it is natural to connect the NLI to the shear component of the tube-orientation tensor. This interpretation is also consistent with molecular tube-model calculations showing that nonlinear oscillatory shear responses of entangled polymers are governed primarily by the deformation and relaxation of tube orientation \cite{Read2011}.

To leading order, we therefore assume that the nonlinear index is proportional to the degree of tube orientation,
\begin{equation}
\mathrm{NLI}\propto S_{xy}.
\end{equation}

The proportionality constant is fixed by the limiting orientational state. For a unit orientation tensor, the maximum admissible shear component is $S_{xy}^{\max}=1/2$, corresponding to complete alignment of tube segments in the shear plane. Requiring that this limiting orientational state maps onto the maximum nonlinear response predicted by the present framework leads to the normalization
\begin{equation}
\mathrm{NLI}\simeq 6S_{xy}^{CL},
\label{eq:NLI_Sxy}
\end{equation}
which immediately yields
\begin{equation}
\mathrm{NLI}_{\max}
=
6S_{xy}^{\max}
=
6\left(\frac{1}{2}\right)
=
3.
\end{equation}

The value $\mathrm{NLI}_{\max}=3$ should therefore be viewed as a consequence of the finite orientational capacity of the tube network together with the adopted normalization of the nonlinear index. In this interpretation, the NLI plays the role of a Fourier-resolved measure of dynamic nonaffinity, quantifying the progressive loss of coherent load-bearing orientation under oscillatory deformation.

Combining \eqref{eq:Sxymax} and \eqref{eq:NLI_Sxy} gives the upper bound
\begin{equation}
\label{eq:NLImax}
    \mathrm{NLI}_{\max}=3.
\end{equation}
Thus an approximately proportional law $\mathrm{NLI}\propto\gamma_0$ may hold over an intermediate nonlinear regime, but it cannot remain valid indefinitely.

\section{CCR, tube dilation, and the constraint-limited orientation closure}

Convective constraint release (CCR) was introduced by Marrucci as a mechanism for flow-induced renewal of topological constraints and was subsequently incorporated into Doi--Edwards-type constitutive descriptions~\cite{Marrucci1996,IannirubertoMarrucci1996}. In oscillatory shear, a scalar CCR-modified relaxation rate may be written as
\begin{equation}
\label{eq:taueff_base}
    \frac{1}{\tau_{\mathrm{eff}}}=\frac{1}{\tau_d(\lambda)}+\beta |\dot{\gamma}|\,|S_{xy}|,
\end{equation}
where $\tau_d$ is the disengagement time, $\lambda$ is the chain stretch ratio defined as the ratio between the instantaneous primitive-path contour length $L$ and its equilibrium value $L_0$. The first term represents the intrinsic tube-survival rate, whereas the second term describes deformation-assisted renewal of topological constraints through the convective motion of oriented primitive-path segments \cite{IannirubertoMarrucciCCR2014}. The quantity $\lambda=1$ corresponds to an unstretched chain, whereas $\lambda>1$ represents chain stretch under flow. Also, $\beta$ is the CCR efficiency. The absolute values ensure that the relaxation rate remains positive during both half-cycles.

For sinusoidal shear,
\begin{equation}
    \dot{\gamma}(t)=\Omega\gamma_0\cos(\Omega t).
\end{equation}
With tube dilation,
\begin{equation}
    a=a_0\lambda^{1/2},\qquad Z=Z_0\lambda^{-1},
\end{equation}
where \(Z_0\) is the equilibrium number of entanglements per chain. Since the reptation time scales as
\begin{equation}
    \tau_d=\tau_e Z^3,
\end{equation}
where $\tau_e$ is the entanglement time, i.e. the characteristic relaxation time associated with motion over an entanglement strand, and $Z$ is the mean number of entanglements per chain. Hence, one obtains
\begin{equation}
    \tau_d(\lambda)=\tau_{d0}\lambda^{-3},
    \qquad
    \frac{1}{\tau_d(\lambda)}=\frac{\lambda^3}{\tau_{d0}},
\end{equation}
with $\tau_{d0}=\tau_e Z_0^3$. The effective nonlinear relaxation rate therefore becomes
\begin{equation}
\label{eq:taueff_dilated}
    \frac{1}{\tau_{\mathrm{eff}}}
    =
    \frac{\lambda^3}{\tau_{d0}}
    +
    \beta |\dot{\gamma}|\,|S_{xy}|.
\end{equation}
The interchain-pressure concept provides a molecular basis for coupling stretch, orientation, and tube modification, because a strongly stretched chain does not relax inside an unchanged equilibrium tube \cite{MarrucciIanniruberto2004}.

The key modification proposed here is to replace the unbounded shear-orientation amplitude by a bounded, constraint-limited component,
\begin{equation}
\label{eq:SxyCL}
    S_{xy}^{CL}(\gamma_0)
    =
    \frac{1}{2}
    \frac{\gamma_0/\gamma_s}{1+\gamma_0/\gamma_s}
    =
    \frac{1}{2}
    \frac{\gamma_0}{\gamma_s+\gamma_0},
\end{equation}
where $\gamma_s$ is the characteristic half-saturation strain.
Its molecular origin will be derived in the next section.

This form gives
\begin{equation}
    S_{xy}^{CL}\simeq
    \frac{1}{2}\frac{\gamma_0}{\gamma_s},
    \qquad
    \gamma_0\ll\gamma_s,
\end{equation}
and
\begin{equation}
    S_{xy}^{CL}\rightarrow \frac{1}{2},
    \qquad
    \gamma_0\gg\gamma_s.
\end{equation}
Thus the physical orientation limit is approached asymptotically rather than imposed through a hard cutoff. This should be interpreted as an amplitude-level closure for the shear-orientation envelope, not as an instantaneous tensor evolution equation.

With this closure, \eqref{eq:taueff_dilated} becomes
\begin{equation}
\label{eq:taueff_CL}
    \frac{1}{\tau_{\mathrm{eff}}^{CL}}
    =
    \frac{\lambda^3}{\tau_{d0}}
    +
    \beta |\dot{\gamma}|\,|S_{xy}^{CL}|.
\end{equation}
At very high deformation, \(S_{xy}^{CL}\rightarrow1/2\), and therefore
\begin{equation}
    \frac{1}{\tau_{\mathrm{eff}}^{CL}}
    \rightarrow
    \frac{\lambda^3}{\tau_{d0}}
    +
    \frac{\beta}{2}|\dot{\gamma}|.
\end{equation}
At the maximum shear-rate envelope \(|\dot{\gamma}|_{\max}=\Omega\gamma_0\),
\begin{equation}
    \frac{1}{\tau_{\mathrm{eff,max}}^{CL}}
    \rightarrow
    \frac{\lambda^3}{\tau_{d0}}
    +
    \frac{\beta}{2}\Omega\gamma_0.
\end{equation}
Thus the relaxation rate continues to increase with \(\gamma_0\) even after the orientation itself is close to its admissible maximum. This is an orientational-capacity limit, not a mobility limit.

\section{Stretch saturation and topological origin of \texorpdfstring{$\gamma_s$}{gamma_s}}
Physically, chain stretch is generated when flow convects and orients the primitive path faster than contour-length fluctuations can relax it. The evolution of the stretch ratio $\lambda$ is therefore governed by a balance between a flow-induced stretching term and a Rouse-type relaxation term, such that the stretch dynamics can be written as
\begin{equation}
\label{eq:stretch}
    \frac{d\lambda}{dt}=\dot{\gamma}S_{xy}^{CL}\lambda-\frac{1}{\tau_R}F_{CL}(\lambda)(\lambda-1),
\end{equation}
where $\tau_R$ is the Rouse time and $F_{CL}$ prevents unphysical stretch beyond the topological limit. A simple form is
\begin{equation}
    F_{CL}(\lambda)=\frac{1}{1-\lambda/\lambda_{\max}}.
\end{equation}
As $\lambda\rightarrow\lambda_{\max}$, $F_{CL}$ diverges and prevents further stretch. The finite stretch limit should therefore be interpreted as a topological limitation rather than a purely geometric finite-extensibility correction. As the effective entanglement density decreases, the tube progressively loses its ability to store additional orientational information, thereby limiting further stretch-induced nonlinear response \cite{ReadJagannathanDas2012}.

The molecular origin of $\lambda_{\max}$ is taken to be the loss of proportional orientation memory when the number of remaining effective entanglements decreases to a small critical value,
\begin{equation}
    Z(\lambda_{\max})=Z_{CL},\qquad Z_{CL}\simeq 2\text{--}3.
\end{equation}
Since tube dilation gives
\begin{equation}
    Z(\lambda)=\frac{Z_0}{\lambda},
\end{equation}
we obtain
\begin{equation}
\label{eq:lambdamax}
    \lambda_{\max}=\frac{Z_0}{Z_{CL}}.
\end{equation}
This result emphasizes that the saturation scale is controlled by the available reservoir of orientational constraints rather than by finite chain extensibility alone.
Thus $\lambda_{\max}$ is not introduced as an arbitrary finite-extensibility parameter; it follows from the assumption that proportional tube-orientation memory is lost once only a few effective constraints remain. This interpretation is consistent with modern tube-model reviews in which polydispersity, long-chain branching, and dynamic dilution modify the effective relaxation spectrum primarily by changing the population of active constraints rather than by altering a single relaxation time \cite{NarimissaWagner2019}.

Before this constraint-limited regime is reached, we use the intermediate nonlinear stretch scaling
\begin{equation}
\label{eq:lambda_scaling}
    \lambda(\gamma_0)\simeq C_\lambda\left(c_\Omega\Omega\tau_d\gamma_0\right)^{1/5},
\end{equation}
where $C_\lambda$ is an order-one prefactor and $c_\Omega$ is a numerical factor of order unity
that depends on whether the peak shear rate \(c_\Omega=1\)
or a cycle-averaged shear rate is used. The characteristic saturation strain $\gamma_s$,
which marks the crossover from approximately proportional
NLI growth to the constraint-limited regime, thus corresponding to the half-saturation point of the
constraint-limited orientation response, 
is defined by
\begin{equation}
    \lambda(\gamma_s)=\lambda_{\max}.
\end{equation}
Within the present crossover model, $\gamma_s$ should not be interpreted as the strain at which saturation is fully reached. Rather, it corresponds to the characteristic half-saturation point of the constraint-limited response. Indeed, from Eq.~(\ref{eq:NLI_curve}),
\begin{equation}
    \mathrm{NLI}(\gamma_s)=\frac{3}{2},
\end{equation}
which is one half of the limiting value $\mathrm{NLI}_{\max}=3$. Full saturation is approached only asymptotically for $\gamma_0\gg\gamma_s$.
Using \eqref{eq:lambdamax} and \eqref{eq:lambda_scaling},
\begin{equation}
\label{eq:gammas}
    \gamma_s=
    \frac{1}{c_\Omega\Omega\tau_{d0}}
    \left(
    \frac{Z_0}{C_\lambda Z_{CL}}
    \right)^5 .
\end{equation}

Since $\tau_d=\tau_e Z_0^3$, this can also be written as
\begin{equation}
\label{eq:gammas_Z0}
    \gamma_s=\frac{Z_0^2}{c_\Omega\Omega\tau_e C_\lambda^5 Z_{CL}^5}.
\end{equation}
For a linear polymer with $Z_0=M/M_e$,
\begin{equation}
\label{eq:gammas_M}
    \gamma_s=\frac{1}{c_\Omega\Omega\tau_e C_\lambda^5 Z_{CL}^5}\left(\frac{M}{M_e}\right)^2.
\end{equation}
Thus, at fixed chemistry and frequency, the theory predicts a quadratic increase of the saturation strain with entanglement number,
\begin{equation}
    \gamma_s \propto Z_0^2.
\end{equation}
More highly entangled linear polymers therefore require larger deformation amplitudes before the orientational constraint reservoir is substantially depleted and the constraint-limited orientation regime is reached.

\section{NLI curve in the constraint-limited regime}

Combining \eqref{eq:NLI_Sxy} and \eqref{eq:SxyCL}, the nonlinear index becomes
\begin{equation}
\label{eq:NLI_curve}
    \mathrm{NLI}(\gamma_0)
    =
    3\,
    \frac{(\gamma_0/\gamma_s)^m}
         {1+(\gamma_0/\gamma_s)^m}.
\end{equation}
The constraint-limited orientation law of
\eqref{eq:SxyCL}
corresponds to the special case $m=1$.
To account phenomenologically for the broader crossover observed in experiments, we introduce an effective crossover exponent $m$, which controls the sharpness of the transition between proportional-growth and saturation regimes without modifying the limiting value $\mathrm{NLI}_{\max}=3$.
For $\gamma_0\ll\gamma_s$,
\begin{equation}
    \mathrm{NLI}(\gamma_0)
    \simeq
    3\left(\frac{\gamma_0}{\gamma_s}\right)^m,
\end{equation}
whereas for $\gamma_0\gg\gamma_s$,
\begin{equation}
    \mathrm{NLI}(\gamma_0)
    \simeq
    3\left[
    1-
    \left(\frac{\gamma_s}{\gamma_0}\right)^m
    \right]
    \rightarrow 3^{-}.
\end{equation}
The exponent $m$ should not be interpreted solely as a fitting parameter. Rather, it provides an effective representation of crossover broadening arising from distributed relaxation times, nonlinear chain stretch, and cycle-dependent constraint renewal \cite{Kapnistos2008}.

\begin{figure}[ht]
\centering
\includegraphics[width=0.82\textwidth]{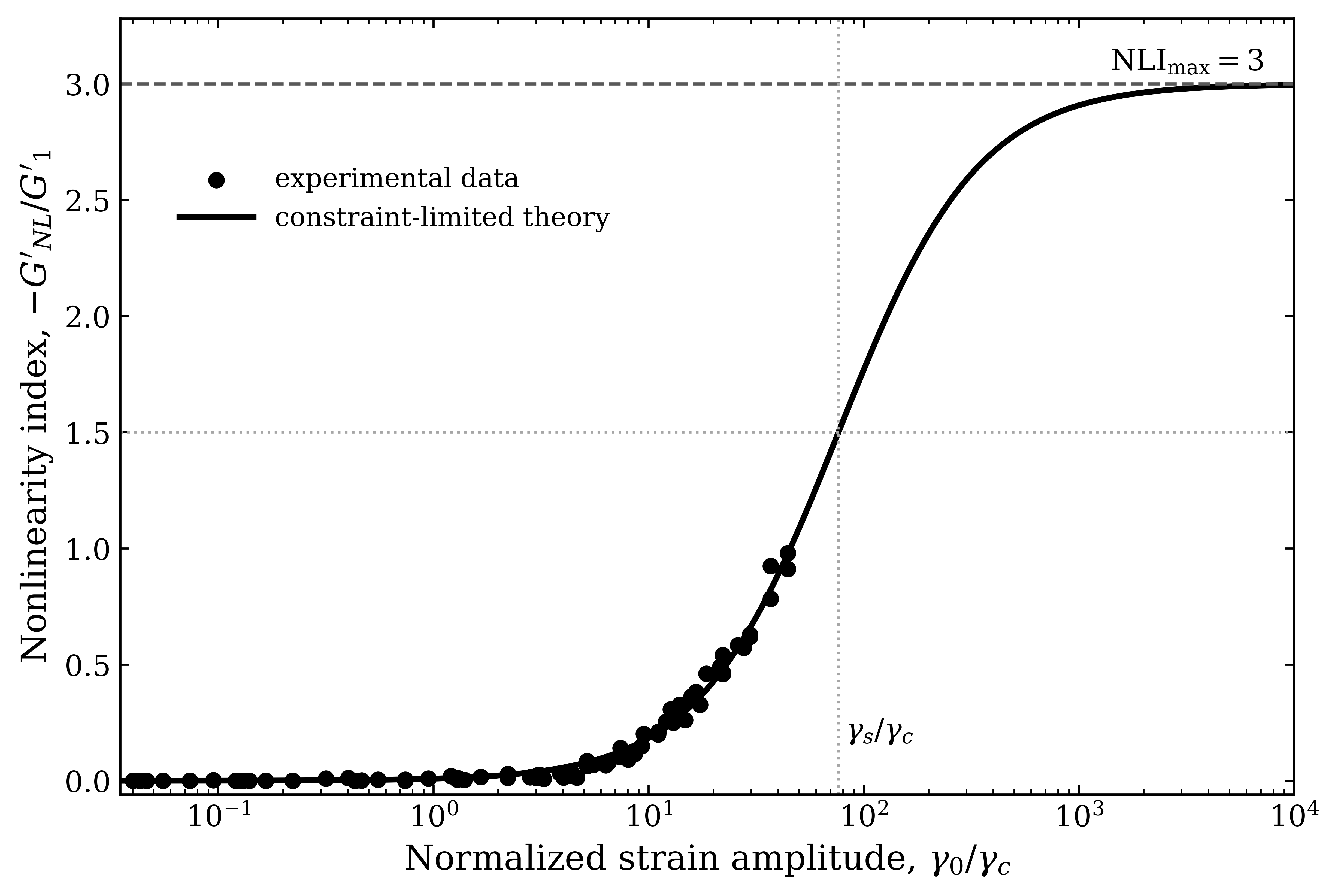}
\caption{
Experimental master curve of the nonlinear index,
$\mathrm{NLI}=-G'_{NL}/G'_1$,
plotted as a function of the normalized strain amplitude
$\gamma_0/\gamma_c$,
where $\gamma_c\equiv\gamma_0^{LR}$ denotes the strain amplitude marking the onset of nonlinear response and departure from linear viscoelasticity. Symbols represent collapsed experimental data obtained from different entangled polymer systems. The solid line is the prediction of the constraint-limited tube-orientation model,
$\mathrm{NLI}
=
3
\frac{(\gamma_0/\gamma_s)^m}
     {1+(\gamma_0/\gamma_s)^m},$
which incorporates the geometrical upper bound $S_{xy}\leq 1/2$ for tube-segment orientation and the corresponding theoretical limiting value $\mathrm{NLI}_{\max}=3$. The fit yields
$\gamma_s/\gamma_c \simeq 76$
and $m\simeq1.34$.
The parameter $\gamma_s$ denotes the characteristic saturation strain at which the nonlinear index reaches one half of its ultimate limiting value,
$\mathrm{NLI}=1.5$,
while full saturation is approached only asymptotically for
$\gamma_0\gg\gamma_s$.
}
\label{fig:nli_master_curve}
\end{figure}

The predicted crossover behavior is shown in \eqref{fig:nli_master_curve}.
The collapse of data from different polymer systems onto a common master curve suggests that the nonlinear index is controlled primarily by the ratio of the imposed strain amplitude to the characteristic nonlinear strain scale, rather than by chemistry-specific details.

The strain scale $\gamma_c$ identifies the transition from the linear viscoelastic regime to the nonlinear regime. It is therefore distinct from the saturation scale $\gamma_s$. The former marks the onset of measurable higher-harmonic generation, whereas the latter controls the subsequent approach toward the constraint-limited plateau. In this sense, $\gamma_c$ is the experimentally accessible onset strain for nonlinear tube dynamics, while $\gamma_s$ characterizes the high-strain orientational-capacity limit.

Molecular architecture primarily affects the value of $\gamma_c$, namely the strain amplitude at which nonlinear harmonics first become measurable. Large changes in molecular weight do not necessarily destroy the onset of linearity, because the normalized strain variable $\gamma_0/\gamma_0^{LR}$ already accounts for the material-dependent departure from linear response. Long-chain branching, however, can delay or modify nonlinear response because branch points act as additional topological constraints and alter the relative relaxation of arms and backbone.

\section{Critical strain for nonlinearity onset}

The critical strain amplitude $\gamma_c$ should be calculated from the first measurable CCR-induced loss of linear orientation memory, not from stretch saturation or from the asymptotic limit $\mathrm{NLI}\rightarrow3$. At $\gamma_c$, the system is just leaving the linear viscoelastic regime, so
\begin{equation}
    \lambda\simeq1,
    \qquad
    Z(\lambda)\simeq Z_0,
    \qquad
    S_{xy}^{CL}\simeq S_{xy}.
\end{equation}
Near onset,
\begin{equation}
    |\dot{\gamma}|\propto\gamma_0,
    \qquad
    |S_{xy}|\propto\gamma_0.
\end{equation}
Therefore, the CCR activation accumulated over one oscillatory cycle scales as
\begin{equation}
    \Phi_{CCR}\propto\gamma_0^2.
\end{equation}
If the molecule has $Z_{\mathrm{eff}}$ active orientational constraints, the number of constraints affected by the first CCR activation can be written as
\begin{equation}
    \Delta Z_{CCR}=C_c Z_{\mathrm{eff}}\gamma_0^2,
\end{equation}
where $C_c$ collects the CCR efficiency, the linear orientation coefficient, and the fixed-frequency response. The onset of nonlinearity is defined by
\begin{equation}
    \Delta Z_{CCR}(\gamma_c)=n_c,
\end{equation}
with $n_c$ an order-one threshold for the first measurable nonlinear harmonic. Hence
\begin{equation}
\label{eq:gamma_c_general}
    \gamma_c=\frac{A_c}{\sqrt{Z_{\mathrm{eff}}}},
    \qquad
    A_c=\left(\frac{n_c}{C_c}\right)^{1/2}.
\end{equation}
At a fixed frequency, $A_c$ may be treated as a common prefactor for comparing polymer architectures.

\section{Architecture dependence of \texorpdfstring{$\gamma_c$}{gamma_c}}

For a linear polymer with no branches,
\begin{equation}
    Z_{\mathrm{eff}}=Z_0=\frac{M}{M_e},
\end{equation}
and therefore
\begin{equation}
\label{eq:gamma_c_linear}
    \gamma_c^{lin}=A_c\left(\frac{M_e}{M}\right)^{1/2}.
\end{equation}
A longer linear chain has more orientational constraints and reaches the first nonlinear onset at a lower strain amplitude.

For a long-chain-branched polymer, define
\begin{equation}
    Z_{bb}=\frac{M_{bb}}{M_e},\qquad Z_{aa}=\frac{M_{aa}}{M_e},
\end{equation}
where $M_{bb}$ is the backbone molecular weight and $M_{aa}$ is the molecular weight of a long arm. If $\nu_a$ is the number of long arms per molecule, a sparse-branching estimate is
\begin{equation}
    Z_{\mathrm{eff}}^{sparse}=Z_{bb}+\eta_a\nu_a Z_{aa},
\end{equation}
where $0<\eta_a\leq1$ is an arm-orientation efficiency factor. Thus
\begin{equation}
\label{eq:gamma_c_sparse}
    \gamma_c^{sparse}=A_c\left[\frac{M_e}{M_{bb}+\eta_a\nu_a M_{aa}}\right]^{1/2}.
\end{equation}
For weak sparse branching, with $\nu_aM_{aa}\ll M_{bb}$,
\begin{equation}
    \gamma_c^{sparse}\simeq A_c\left(\frac{M_e}{M_{bb}}\right)^{1/2}
    \left[1-\frac{1}{2}\eta_a\frac{\nu_aM_{aa}}{M_{bb}}\right].
\end{equation}
Sparse branching therefore gives a small correction to the corresponding linear backbone value.

\begin{figure}[ht]
\centering
\includegraphics[width=0.65\textwidth]{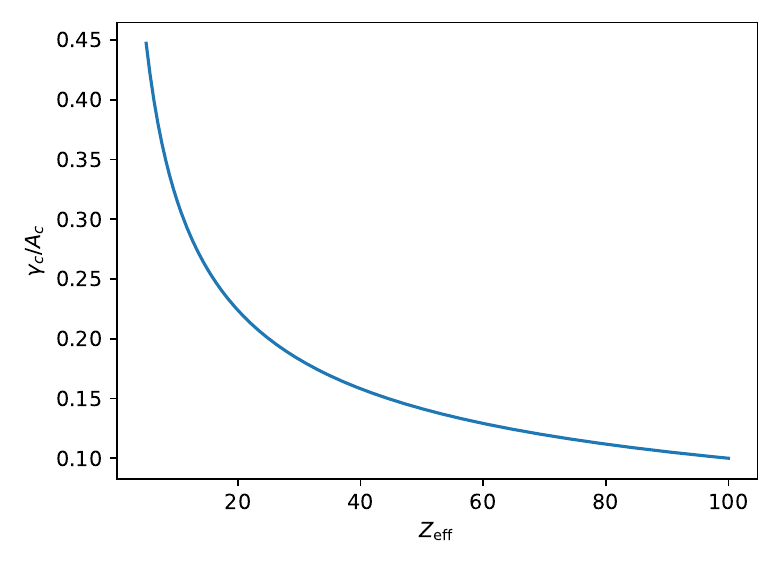}
\caption{General CCR-onset scaling $\gamma_c/A_c=Z_{\mathrm{eff}}^{-1/2}$. Molecular architecture enters through the effective number of orientational constraints $Z_{\mathrm{eff}}$.}
\label{fig:gamma_c_scaling}
\end{figure}

For regular or more densely long-chain-branched polymers, the arm contribution should saturate because arms and branch points become dynamically coupled. A compact saturated form is
\begin{equation}
    Z_{\mathrm{eff}}^{reg}=Z_{bb}\left\{1+\eta_a\left[1-\exp\left(-\frac{\nu_a Z_{aa}}{Z_{bb}}\right)\right]\right\}.
\end{equation}
The corresponding onset strain is
\begin{equation}
\label{eq:gamma_c_regular}
    \gamma_c^{reg}=A_c
    \left[
    \frac{M_e}{M_{bb}\left\{1+\eta_a\left[1-\exp\left(-\frac{\nu_aM_{aa}}{M_{bb}}\right)\right]\right\}}
    \right]^{1/2}.
\end{equation}
The limiting behavior of \eqref{eq:gamma_c_general} is illustrated in \eqref{fig:gamma_c_scaling}.

\section{Physical picture and regimes}

The model distinguishes two strain scales. The first, $\gamma_c$, marks the first measurable departure from linear viscoelasticity and is controlled by the activation of CCR over one cycle. The second, $\gamma_s$, marks the onset of constraint-limited orientation and is controlled by the depletion of the orientable entanglement reservoir down to $Z_{CL}\simeq2$--$3$.

For $\gamma_0<\gamma_c$, the response is mainly first-harmonic and tube orientation memory remains essentially linear. For $\gamma_c<\gamma_0<\gamma_s$, CCR and tube deformation generate higher harmonics, and the NLI may grow approximately proportionally to strain amplitude. For $\gamma_0>\gamma_s$, chain stretch and tube dilation have reduced the effective entanglement count to a small residual value; the shear orientation is close to its geometric upper bound, and the NLI approaches 3 asymptotically.

Linear high-molecular-weight polymers are expected to display a broad proportional-growth regime because $\gamma_s\propto(M/M_e)^2$. Sparse long-chain branching may shift $\gamma_c$ modestly while preserving a linear-like growth window. More regular or dense branching can lower the nonlinear onset but may also introduce additional relaxation pathways, such as arm retraction and branch-point motion, which can suppress or reshape the high-strain NLI response.

Although the present theory provides an explicit molecular prediction for the onset strain $\gamma_c$, the corresponding influence of molecular architecture on the saturation scale $\gamma_s$ remains more speculative and should be studied in detail in future work. In this perspective, ring polymers provide a particularly useful test case because their nonlinear response is controlled by topological constraints and molecular architecture in the absence of free chain ends. As a result, they may offer an attractive benchmark for distinguishing universal tube-orientation effects from architecture-specific relaxation mechanisms and may provide a stringent experimental test of the constraint-limited framework developed here \cite{Kapnistos2008,Rubinstein2011,Parisi2021JRheol,Parisi2021Macromolecules}.

\section{Discussion}

The available experimental data probe only the initial part of the
approach toward the theoretical saturation limit
$\mathrm{NLI}_{\max}=3$.
Although the measurements do not yet extend into the fully saturated
regime, they are consistent with the predicted crossover from
approximately proportional growth to an asymptotic orientational
capacity limit.

The experimental data currently probe only the rising branch of the predicted crossover. Consequently, the value $\mathrm{NLI}_{\max}=3$ should be regarded as a molecular upper bound implied by the orientational-capacity limit of the tube network rather than as a directly measured experimental plateau.

The fitted crossover exponent \(m \simeq 1.34\) is larger than the simplest proportional-growth expectation. Within the present framework this indicates that the growth of higher harmonics is amplified by additional nonlinear mechanisms beyond the first activation of CCR, including tube dilation, chain stretch, and heterogeneous loss of orientational memory during the oscillation cycle.

The model is intentionally formulated in terms of a small number of effective molecular variables. Real commercial polymers may have broad and non-Maxwellian relaxation spectra, multiple tube times, and strong polydispersity. These effects will broaden the crossover and may alter the apparent values of \(N_{\max}\), \(\gamma_c\), and \(m\). The present theory should therefore be viewed as a molecular closure for the dominant orientational nonlinearity, rather than as a complete constitutive model for all relaxation modes.

Finally, the present work focuses on nonlinear elastic harmonic generation and therefore on $G'_1$ and $G'_{NL}$. In polymer melts the loss modulus and dissipative harmonics can be equally important, especially when $G''>G'$. Extending the same constraint-limited framework to the dissipative nonlinear response is a natural direction for future work.

A similar construction can be applied to the viscous contribution. The generalized nonlinear loss modulus may be written as
\begin{equation}
G''_0(\gamma_0)
=
\frac{\sigma''_1}{\gamma_0}
+
\left[
-3\frac{\sigma''_3}{\gamma_0}
+5\frac{\sigma''_5}{\gamma_0}
-7\frac{\sigma''_7}{\gamma_0}
+\cdots
+(-1)^n(2n+1)\frac{\sigma''_{2n+1}}{\gamma_0}
\right].
\end{equation}
The alternating signs follow from the symmetry of the cosine harmonics and encode the phase structure of nonlinear viscous dissipation. Thus, the present framework can in principle be extended from nonlinear elastic storage to nonlinear viscous dissipation. In this work we focus on the elastic NLI, but the corresponding generalized loss modulus provides a natural route for analyzing nonlinear changes in $\tan\delta$, secondary loops, and intra-cycle dissipative inversion phenomena.

The present theory addresses the monotonic regime of NLI growth. In strongly long-chain-branched polymers additional relaxation mechanisms associated with arm retraction, branch-point motion and competitive arm reptation may become dominant before the orientational-capacity limit is reached. In such cases the NLI may exhibit a maximum followed by a decrease at larger strain amplitudes. The description of this regime requires an extension beyond the present constraint-limited framework and will be addressed separately.

\subsection{Experimental predictions}
A particularly stringent test of the theory would be provided by
monodisperse linear polymers spanning a broad range of entanglement
numbers, for which both $\gamma_c$ and $\gamma_s$ can be independently
determined from Fourier harmonic analysis.
The present framework leads to several experimentally testable predictions.

First, the theory predicts a universal upper bound for the nonlinear index,
\begin{equation}
\mathrm{NLI}_{\max}=3,
\end{equation}
which follows from the geometrical limit $S_{xy}\leq 1/2$ for tube-segment orientation. This prediction is independent of molecular weight, chemistry, and architecture, provided that the response remains dominated by the monotonic constraint-limited regime considered here.

Second, the onset strain for nonlinear behavior is predicted to scale with the effective density of orientational constraints according to
\begin{equation}
\gamma_c \propto Z_{\mathrm{eff}}^{-1/2},
\end{equation}
implying that more highly entangled systems enter the nonlinear regime at smaller strain amplitudes.

Third, for linear polymers the characteristic saturation strain is predicted to scale as
\begin{equation}
\gamma_s
\propto
\left(\frac{M}{M_e}\right)^2,
\end{equation}
indicating that longer chains possess a larger reservoir of orientational constraints and therefore require larger deformations before approaching the orientational-capacity limit.

Finally, the theory predicts a separation between the strain scale associated with the onset of nonlinearity and the strain scale associated with orientational saturation,
\begin{equation}
\gamma_c < \gamma_s.
\end{equation}
The former controls the appearance of measurable higher harmonics, whereas the latter governs the subsequent crossover toward the asymptotic nonlinear limit.

\section{Conclusions}

We have developed a CCR-based molecular closure for the nonlinear index in entangled polymers under periodic shear. The theory is based on the geometrical upper bound of the shear component of the tube-orientation tensor, $S_{xy}\leq1/2$, which leads naturally to the limiting value $\mathrm{NLI}_{\max}=3$ in the normalization used here. The constraint-limited orientation law
\begin{equation}
    S_{xy}^{CL}(\gamma_0)=\frac{1}{2}\frac{\gamma_0}{\gamma_s+\gamma_0}
\end{equation}
provides the simplest smooth interpolation between proportional NLI growth and asymptotic saturation. Experimental master-curve data are described more accurately by a generalized crossover form containing an effective exponent $m$, which accounts phenomenologically for crossover broadening while preserving the same limiting value $\mathrm{NLI}_{\max}=3$.

The saturation strain $\gamma_s$ is not treated as a purely empirical cutoff. It is connected to the loss of orientation memory when the number of remaining effective entanglements decreases to $Z_{CL}\simeq2$--$3$. This gives
\begin{equation}
    \gamma_s=\frac{1}{c_\Omega\Omega\tau_{d0}}\left(\frac{Z_0}{C_\lambda Z_{CL}}\right)^5,
\end{equation}
or, for linear polymers in the ideal reptation scaling limit, $\gamma_s\propto(M/M_e)^2$.

The nonlinear onset strain $\gamma_c$ is controlled by the first CCR-induced loss of linear orientation memory and scales as $\gamma_c=A_cZ_{\mathrm{eff}}^{-1/2}$. This provides explicit predictions for linear, sparsely long-chain-branched, and more regularly branched polymers. Overall, the framework provides a physically transparent
connection between Fourier harmonic observables under periodic shear,
CCR, tube dilation, chain stretch, and the finite orientational
capacity of the entangled tube network.
The theory predicts two distinct nonlinear strain scales:
the onset strain $\gamma_c$ associated with the first measurable CCR-induced loss of orientational memory,
and the half-saturation strain $\gamma_s$ associated with exhaustion of the orientational constraint reservoir.

The present framework is intended as a physically transparent
description of how finite orientational capacity, convective
constraint release, and tube dilation combine to shape the nonlinear
response of entangled polymers under periodic shear. While the
quantitative details may depend on molecular architecture and
relaxation-pathway competition, the existence of distinct onset
$\gamma_c$ and saturation $\gamma_s$ strain scales appears to be a
robust consequence of the finite orientational memory of the
entanglement network.

\section*{Declarations}

\subsection*{Conflict of interest}

The authors declare that they have no conflict of interest.

\subsection*{Author contributions}

D. Nichetti developed the nonlinear rheological framework,
performed the data analysis, and contributed to the theoretical
development. A. Zaccone developed the nonaffine and molecular-theory
interpretation, supervised the theoretical work, and contributed to
manuscript preparation. Both authors discussed the results and approved
the final manuscript.

\subsection*{Data availability}

The data supporting the findings of this study are available from the
corresponding author upon reasonable request.

\subsection*{Code availability}

Not applicable.

\bibliographystyle{unsrt}
\bibliography{references}

\end{document}